\documentclass[11pt]{article}

\usepackage{graphicx}
\usepackage{amssymb}
\usepackage{amsmath}
\usepackage{amsfonts}
\usepackage{booktabs}
\usepackage[round]{natbib}
\usepackage{caption}
\usepackage{subcaption}
\usepackage[flushleft]{threeparttable}
\usepackage{graphicx}
\usepackage{bm}
\usepackage{multirow}
\usepackage{color}
\usepackage{fullpage}
\usepackage{setspace}

\newtheorem{lem}{Lemma}

\newtheorem{lemma}{Lemma}

\usepackage{color}

\DeclareSymbolFont{AMSb}{U}{msb}{m}{n}
\DeclareMathSymbol{\N}{\mathbin}{AMSb}{"4E}
\DeclareMathSymbol{\Z}{\mathbin}{AMSb}{"5A}
\DeclareMathSymbol{\R}{\mathbin}{AMSb}{"52}
\DeclareMathSymbol{\Q}{\mathbin}{AMSb}{"51}
\DeclareMathSymbol{\I}{\mathbin}{AMSb}{"49}
\DeclareMathSymbol{\C}{\mathbin}{AMSb}{"43}
\DeclareMathSymbol{\D}{\mathbin}{AMSb}{"44}
\DeclareMathSymbol{\E}{\mathbin}{AMSb}{"45}

\def\b1{\boldsymbol{1}}
\def\RR{\hbox{I\kern-.1667em\hbox{R}}}

\newcommand{\EE}[1]{\E \left [ #1 \right ]}

\newcommand{\ind}{\perp \! \! \! \perp}
\newcommand{\obs}{\text{obs}}

\newcommand{\subtitle}[1]{%
  \posttitle{%
    \par\end{center}
    \begin{center}\large#1\end{center}
    \vskip0.5em}%
}

\begin{document}

\title{Principal Score Methods:\\Assumptions and Extensions\thanks{\noindent{\scriptsize\textit{Email: } \texttt{afeller@berkeley.edu}. AF and LM gratefully acknowledge financial support from the Spencer Foundation through a grant entitled ``Using Emerging Methods with Existing Data from Multi-site Trials to Learn About and From Variation in Educational Program Effects,'' and from the Institute for Education Science (IES Grant \#R305D150040). We would like to thank Alberto Abadie, Peng Ding, Jennifer Hill, Jiannan Lu, Don Rubin, Elizabeth Stuart, and members of the Spencer group for helpful comments and discussion, as well as seminar participants at the 2015 Society for Research on Educational Effectiveness meeting. All opinions expressed in the paper and any errors that it might contain are solely the responsibility of the authors.}}}
\author{Avi Feller\\UC Berkeley \and Fabrizia Mealli\\Universit{\`a} di Firenze \and Luke Miratrix\\Harvard GSE}
\maketitle

\begin{abstract}
Researchers addressing post-treatment complications in randomized trials often turn to principal stratification to define relevant assumptions and quantities of interest. One approach for estimating causal effects in this framework is to use methods based on the ``principal score,'' typically assuming that stratum membership is as-good-as-randomly assigned given a set of covariates. In this paper, we clarify the key assumption in this context, known as Principal Ignorability, and argue that versions of this assumption are quite strong in practice. We describe different estimation approaches and demonstrate that weighting-based methods are generally preferable to subgroup-based approaches that discretize the principal score. We then extend these ideas to the case of two-sided noncompliance and propose a natural framework for combining Principal Ignorability with exclusion restrictions and other assumptions. Finally, we apply these ideas to the Head Start Impact Study, a large-scale randomized evaluation of the Head Start program. Overall, we argue that, while principal score methods are useful tools, applied researchers should fully understand the relevant assumptions when using them in practice. 
\end{abstract}

\onehalfspacing

\section{Introduction}

Although the principal stratification framework has gained widespread use for defining estimands of interest~\citep[for a recent review, see,][]{Page2015}, the method of estimation can differ dramatically from application to application. 
In the article first defining principal stratification, for example,~\citet{Frangakis:2002wp} advocate the use of a full model-based estimation strategy, such as that found in~\citet{Imbens:1997bayes} and~\citet{Hirano:2000vg}. While this strategy is relatively common in statistics and biostatistics, there has been limited adoption of this approach among education and policy researchers, perhaps due to the complexity of implementation and unfamiliarity with Bayesian and likelihood methods.

In this paper, we explore an alternative approach that leverages covariates and various conditional independence assumptions to identify target estimands of interest. In particular, we address \textit{principal score methods}, which rely on predictive covariates (rather than outcome distributions) for estimating principal causal effects. The goal of this paper is to review the existing methods and to clarify the assumptions necessary for the proposed procedures to yield estimates of the causal estimands of interest~\citep[see also][for recent discussion]{Ding:2014wc}. 
We investigate the role of the Principal Ignorability assumption in this approach, show how there are in fact two main versions, which we term Strong and Weak Principal Ignorability, and compare these assumptions to other ignorability assumptions in the literature. We then explore two estimation methods proposed for the one-sided case, the principal score weighting method first proposed by~\citet{Jo:2009hx} and the discrete subgroup method of~\citet{Schochet:2007ky}. We show that the weighting method only requires Weak Principal Ignorability in this setting. By contrast, the discrete subgroup method does not appear to unbiasedly estimate the principal causal effects of interest even under Strong Principal Ignorability. We confirm this result with simulation studies.

Next, we explore the more complex case of two-sided noncompliance. 
We demonstrate how researchers can ``mix'' principal ignorability assumptions with more common assumptions from causal inference, namely exclusion restrictions. We then apply these methods to the Head Start Impact Study~\citep{puma2010hsis}, a large-scale randomized evaluation of the Head Start program, finding mixed results overall.

Overall, we believe that this is a useful contribution to the small-but-growing literature on principal score methods. Like many statistical concepts, the idea of the principal score has multiple origins in different sub-fields. In biostatistics, the concept was first formalized by~\citet{Follmann:2000fm}, who called this the \textit{compliance score}~\citep[see also][]{Joffe:2003ic, Aronow:2013et}. In the literature on statistics in the social sciences, the idea is due to~\citet{Hill:2002dt}, who introduced the term \textit{principal score}~\citep[see also][]{Jo:2002cw,Jo:2009hx,Stuart:2011gd}. There have been many examples of this approach in practice, particularly in education and program evaluation, with some recent prominent examples from~\citet{Schochet:2007ky} and~\citet{Zhai:2014ca}.~\citet{Schochet:2014wa} offer a recent overview.~\citet{Porcher:2015bi} give a recent simulation study.~\citet{Ding:2014wc} give theoretical justification for a more general setup and offer additional guidance on estimation and sensitivity analysis. 
This paper could be considered a helpful applied complement to the recent work of~\citet{Ding:2014wc}.

The paper proceeds as follows. 
Section~\ref{sec:one_sided_estimands} defines the relevant estimands and assumptions in the case of one-sided noncompliance.
Section~\ref{sec:one_sided_estimation} discusses estimation in this setting.
Section~\ref{sec:two_sided} extends these ideas to the case of two-sided noncompliance.
Section~\ref{sec:hsis} applies the underlying methods to the Head Start Impact Study.
Section~\ref{sec:discussion} offers some thoughts for future research and concludes. The Appendix includes a short proof of some desirable properties of the principal score.

\section{Estimands and assumptions in the one-sided case\label{sec:one_sided_estimands}}
\subsection{Setup}
We begin with a simple toy example of a supplemental tutoring program in a school. We observe a total of $N$ students, $N_1$ of whom are randomized to receive this supplemental program, with treatment indicator $Z_i = 1$ for student $i$, and $N_0$ of whom are not, with $Z_i = 0$. For ease of exposition, we assume that random assignment is via complete randomization and invoke SUTVA~\citep{imbens_rubin2014}.

In this toy example, some of the students assigned to treatment receive a ``high'' dose of the program while some receive a ``low'' dose. For example, all students assigned to treatment attend one mandatory tutoring session each week, but some students can also decide to attend an optional second session each week. Students who only attend the required one weekly session receive the low dose; students who also attend the second session receive the high dose. The complication is that unobserved factors determine whether students assigned to treatment attend one versus two weekly sessions. Formally, let $D_i(1) \in \{L, H\}$ denote whether student $i$ would receive a low or high dose of the intervention if assigned to $Z_i = 1$. 
In this example all students have $D_i(0) = 0$ for no tutoring at all.

The above scenario corresponds to \textit{one-sided noncompliance}: students not assigned to treatment have no access to any tutoring. In the classic noncompliance setting, ``low dose'' would correspond to not taking the treatment when offered (i.e., Never Takers). Since the corresponding estimands are more interesting here, we instead consider the case where students assigned to treatment could receive either of two levels of treatment. 

Following~\citet{angrist1996identification} and~\citet{Frangakis:2002wp}, we define \textit{compliance types} or \textit{principal strata} based on the joint values of treatment received under treatment and control, $(D_i(0), D_i(1))$. Since $D_i(0) = 0$ for all students, principal strata are completely defined by $D(1)$~\citep{imbens_rubin2014}. 
With some abuse of terminology, we define these two types as:

$$ S_i \equiv \begin{cases}
	\text{Low Takers ($\ell$)} & \text{if } D_i(0) = 0 \text{ and } D_i(1) = L \\
	\text{High Takers (h)} & \text{if } D_i(0) = 0 \text{ and } D_i(1) = H,
	\end{cases}$$
where Low Taker here means a student who never takes a high dose of the program, and where $S_i = \text{$\ell$}$ and $S_i = \text{h}$ indicate that individual $i$ is a Low Taker or High Taker, respectively.

Importantly, since we regard potential outcomes as fixed, the joint values $(D_i(0), D_i(1))$ are also fixed for each individual, and we can regard $S$ as a pre-treatment covariate. Therefore, we can think of subgroup treatment effects for Low  or High Takers the same way we would consider subgroup effects among men and women. With this in mind, we are interested in the separate effects of receiving a low dose of the program (versus no dose) and of receiving a high dose of the program (versus no dose). Within each principal stratum, it is as if we have a randomized experiment that could allow us to estimate these \textit{principal causal effects} of interest~\citep{Frangakis:2002wp}:\footnote{Note that this paper focuses on super-population estimands, which appear to be the objects of interest in the principal score literature. We are not aware of any discussion of finite sample versus super-population estimands in this setting. See~\citet{imbens_rubin2014} for further discussion of finite versus super-population inference.}
\begin{align*}
\text{ITT}_{\text{h}} &= \mathbb{E}\{Y_i(1) - Y_i(0) \mid S_i = \text{h}\} = \mu_{\text{h}1} - \mu_{\text{h}0}, \\
\text{ITT}_{\text{$\ell$}} &= \mathbb{E}\{Y_i(1) - Y_i(0) \mid S_i = \text{$\ell$}\} = \mu_{\text{$\ell$}1} - \mu_{\text{$\ell$}0},
\end{align*}
where $\mu_{sz} \equiv \mathbb{E}\{Y_i(z) \mid S_i = s\}$.

One major benefit of randomization is that pre-treatment characteristics are balanced across treatment conditions on average. We can therefore obtain unbiased estimates of key population quantities via their distribution in only one treatment condition. In particular, because we directly observe stratum membership for those individuals assigned to treatment we can immediately estimate $\pi \equiv \mathbb{P}\{S_i = \text{h}\}$, $\mu_{\text{h}1} \equiv \mathbb{E}\{Y_i(1) \mid S_i = \text{h}\}$, and $\mu_{\text{$\ell$}1} \equiv \mathbb{E}\{Y_i(1) \mid S_i = \text{$\ell$}\}$. 

This is not the case for similar quantities among those individuals assigned to control. For this group, we observe a mixture of types, since $D_i(0) = 0$ for all $i$, but cannot directly observe who would be a High Taker or a Low Taker.
Therefore, we cannot immediately estimate $\mu_{\text{h}0}$ and  $\mu_{\text{$\ell$}0}$, and, as a result, cannot estimate $\text{ITT}_{\text{h}}$ and $\text{ITT}_{\text{$\ell$}}$.
Table~\ref{tbl:onesided_example} shows this mixture problem.

	\begin{table}[btp]
	\centering
	\begin{threeparttable}
		\centering
			\caption{\label{tbl:onesided_example} Possible principal strata in one-sided toy example}
	\begin{tabular}{ccl}
			\toprule
			$Z_i$ & $D_i^{\text{ obs}}$ &  \multicolumn{1}{c}{\textbf{Possible principal strata}}\\
			\midrule
			$1$ & H & High Taker (treatment) \\[0.5em]
			$1$ & L & Low Taker (treatment) \\[0.5em]
			$0$ & $0$ & Low Taker (control); High Taker (control)\\
			\bottomrule
		\end{tabular}
		\end{threeparttable}
	\end{table}

Following~\citet{angrist1996identification}, we could avoid these estimation challenges by assuming the exclusion restriction for the Never Takers, that is, by assuming that $\text{ITT}_{\text{$\ell$}} = 0$. In our example, this assumes that there is no impact of receiving a low dose of the program (versus receiving no dose). If this assumption is incorrect, the resulting estimate for $\text{ITT}_{\text{h}}$ could biased.
In particular, it could be greatly overstated because we allocate the full effect of the overall treatment to the high dose group only.

Before turning to principal score methods, we note that there are a range of alternative approaches that broadly fall under the umbrella of principal stratification. 
One option is to use a fully model-based estimation strategy, such as originally proposed by~\citet{Imbens:1997bayes}, which requires imposing distributional assumptions on the outcome to disentangle the mixture.
Alternatively, we could use non-parametric bounds~\citep[e.g.,][]{Zhang:2003kk}, potentially sharpening them by leveraging pre-treatment covariates~\citep{Grilli:2008ih,Long:2013hs} or secondary outcomes~\citep{Mealli:2013be}. 
Such bounds, even with these additional restrictions, can often be too wide for practical use. 
Finally, we could exploit specific conditional independence assumptions between covariates and outcomes conditional on principal strata~\citep{ding2012identifiability, mealli2016concentration} or between outcomes conditional on principal strata~\citep{mealli2016concentration, Mealli:2013be, mattei2013} to achieve full identification of principal causal effects.

\subsection{Principal Ignorability in the one-sided case}
For each student, $i$, we observe a vector of pre-treatment covariates, denoted $\mathbf{x}_i$. The question is: given these covariates, under what assumptions can we obtain reasonable estimates of the causal estimands of interest, $\text{ITT}_{\text{h}}$ and $\text{ITT}_{\text{$\ell$}}$? The key insight is to borrow ideas from the decades of research on using propensity score methods to estimate causal effects in observational studies. Analogous to the propensity score case, the critical assumption we use is that, conditional on covariates, stratum membership is ignorable. Following~\citet{Jo:2009hx}, we call this assumption \textit{Principal Ignorability} (PI). Importantly, we clarify that this assumption has two main forms: \textit{Strong Principal Ignorability} and \textit{Weak Principal Ignorability}. See~\citet{Ding:2014wc} for additional discussion of these assumptions. Finally, in Section~\ref{sec:pscore_ignorability_comparison}, we compare these assumptions to two closely related assumptions in the literature: ignorability and sequential ignorability.

\subsubsection{Strong Principal Ignorability}
In the one-sided case, the \textit{Strong Principal Ignorability} assumption is the pair of conditional independence assumptions:
\begin{eqnarray*}
Y_i(1) &\ind& D_i(1) \mid \mathbf{X}_i, \\
Y_i(0) &\ind& D_i(1) \mid \mathbf{X}_i .
\end{eqnarray*}
This very strong assumption states that, conditional on observed covariates,  whether a student receives a low dose or high dose of the program is unrelated either to that student's outcome when offered the program or that student's outcome in the absence of the program. 
In other words, subgroups defined by pre-treatment covariates can entirely explain any treatment effect variation related to program dosage. 
In particular, knowledge of what dose a student received gives no additional information on their outcomes given their covariates.
This assumption, used in~\citet{Jo:2009hx} and~\citet{Stuart:2011gd}, seems quite strong.

It is instructive to re-write these assumptions in terms of mean independence rather than full stochastic independence (i.e., $\ind$):\footnote{While mean independence is technically weaker than full stochastic independence, it is difficult to imagine a real-world situation in which PI holds in terms of mean independence but not full stochastic independence. For clarity we use the mean formulation.}
\begin{align*}
\mathbb{E}[Y_i(1) \mid \mathbf{X}_i = \mathbf{x}, D_i(1) = 1] &\quad=\quad \mathbb{E}[Y_i(1) \mid \mathbf{X}_i = \mathbf{x}, D_i(1) = 0] \quad=\quad \mathbb{E}[Y_i(1) \mid \mathbf{X}_i = \mathbf{x}], \\
\mathbb{E}[Y_i(0) \mid \mathbf{X}_i = \mathbf{x}, D_i(1) = 1] &\quad=\quad \mathbb{E}[Y_i(0) \mid \mathbf{X}_i = \mathbf{x}, D_i(1) = 0] \quad=\quad \mathbb{E}[Y_i(0) \mid \mathbf{X}_i = \mathbf{x}].
\end{align*}
The above says a student with a given $\mathbf{X}_i$ who received a high dose would have the same expected outcome as a student with that same $\mathbf{X}_i$ who received a low dose.
Again, knowledge of dose does not change the predicted outcome given covariate $\mathbf{X}_i$.

Since this is a randomized experiment, $\mathbb{E}[Y_i(z) \mid \mathbf{X}_i = \mathbf{x}] = \mathbb{E}[Y_i^{\text{obs}} \mid \mathbf{X}_i = \mathbf{x}, Z_i = z]$. Hence, we can write these equalities more compactly as
\begin{align*}
	\mu_{\text{h}1 \mid \mathbf{x}} &= \mu_{\text{$\ell$}1 \mid \mathbf{x}} = \mu_{1\mid \mathbf{x}},\\ 
	\mu_{\text{h}0 \mid \mathbf{x}} &= \mu_{\text{$\ell$}0 \mid \mathbf{x}} = \mu_{0\mid \mathbf{x}},
\end{align*}
where $\mu_{sz \mid \mathbf{x}} \equiv \mathbb{E}\{Y_i(z) \mid S_i = s, \mathbf{X}_i = \mathbf{x}\}$ and $\mu_{z\mid \mathbf{x}} \equiv \mathbb{E}\{Y_i^{\obs} \mid Z_i = z, \mathbf{X}_i = \mathbf{x}\} = \mathbb{E}\{Y_i(z) \mid  \mathbf{X}_i = \mathbf{x}\}$.

\subsubsection{Weak Principal Ignorability}
As shown in Table~\ref{tbl:onesided_example}, we directly observe stratum membership for those students assigned to treatment in this simple example. 
As a result, the first assumption, $Y_i(1) \ind D_i(1) \mid \mathbf{X}_i$, is unnecessary: we can simply estimate the relevant mean outcomes. 
In fact, as we discuss below, we can compare these direct estimates to what we should get if this assumption were true, giving an immediate testable implication.

More naturally, we can relax our strong ignorability assumptions from the pair of assumptions to the single assumption of \textit{Weak Principal Ignorability}, $Y_i(0) \ind D_i(1) \mid \mathbf{X}_i$, which we express as
$$\mu_{\text{h}0 \mid \mathbf{x}} = \mu_{\text{$\ell$}0 \mid \mathbf{x}} = \mu_{0\mid \mathbf{x}} . $$

In words, given covariates $X_i$, we expect the same outcome without intervention for both a Low Taker and a High Taker.
This expected outcome is therefore the same as the observed mean outcome of the mixture of Low and High Takers with that value of $X_i$.
While strictly weaker than Strong PI, Weak PI is not necessarily a weak assumption; it could be difficult to justify in practice.

\subsection{\label{sec:pscore_ignorability_comparison}Comparison with other ignorability assumptions}
We offer a brief comparison of these assumptions to two other assumptions common in the literature: ignorability and sequential ignorability. These approaches have closer ties to classic observational study methods such as matching or propensity scores, in which the researcher models an assignment mechanism and then estimates effects based on that mechanism. Importantly, these alternative paths allow the researcher to estimate effects for the entire population. While appealing, this comes at a cost: we must be able to imagine a hypothetical experiment in which $D$ could plausibly be assigned at random~\citep[e.g.,][]{Rubin:2005hf}. In this case, the principal stratification framework generally requires weaker assumptions but also restricts attention to more local quantities of interest.

\paragraph{Ignorability of $D$.} If we conceive of $D$ as if it were randomly assigned, given covariates, and define potential outcomes $Y_i(D_i = \text{H})$, $Y_i(D_i = \text{L})$, and $Y_i(D_i = \text{0})$, we can think of our units with a specific value of $X_i$ as having been randomized into one of three levels of treatment. 
This is captured by the classic Ignorability assumption:
\begin{align*}
Y_i(D_i = \text{H}) &\ind D_i \mid \mathbf{X}_i,\\
Y_i(D_i = \text{L}) &\ind D_i \mid \mathbf{X}_i,\\
Y_i(D_i = \text{0}) &\ind D_i \mid \mathbf{X}_i.
\end{align*}
In words, this assumes that, given covariates, whether a student receives no dose, a low dose, or a high dose of the program is as good as random. Since potential outcomes are only defined in terms of $D$ rather than $Z$, this framing of the problem does not include any information about the randomization itself. 
The analogous estimands to $\text{ITT}_{\text{h}}$ and $\text{ITT}_{\text{$\ell$}}$ are therefore:
\begin{align*}
\text{ITT}^{\text{Ign}}_{\text{h}} &= \mathbb{E}\{Y_i(D_i = \text{H}) - Y_i(D_i = 0)\}, \\
\text{ITT}^{\text{Ign}}_{\text{$\ell$}} &= \mathbb{E}\{Y_i(D_i = \text{L}) - Y_i(D_i = 0)\}.
\end{align*}
Importantly, these estimands are defined for the entire (super) population of students, rather than for a specific principal stratum.

\paragraph{Sequential Ignorability of $D$.} The Sequential Ignorability assumption for $D$ concieves of \textit{both} $Z$ and $D$ as if they could be randomly assigned, as in a two-stage randomization scheme or factorial design. 
Under this formulation we doubly index the potential outcomes as $Y_i(z,d)$, leading to six possible combinations: $Y_i(1, \text{H})$, $Y_i(1, \text{L})$, $Y_i(1, 0)$, $Y_i(0, \text{H})$, $Y_i(0, \text{L})$, and $Y_i(0, 0)$. 
We then state the Sequential Ignorability assumption as:
$$ Y_i(z,d) \ind D_i \mid \mathbf{X}_i, Z_i \; \mbox{ for } z \in \{0,1\} \mbox{ and } d \in \{0, \text{L}, \text{H} \}.$$
The analogous estimands to $\text{ITT}_{\text{h}}$ and $\text{ITT}_{\text{$\ell$}}$ are therefore:
\begin{align*}
\text{ITT}^{\text{SI}}_{\text{h}} &= \mathbb{E}\{Y_i(1, \text{H}) - Y_i(0, 0)\} \\
\text{ITT}^{\text{SI}}_{\text{$\ell$}} &= \mathbb{E}\{Y_i(1, \text{L}) - Y_i(0, 0)\}.
\end{align*}
As in the typical ignorability case, this estimand is defined for the entire (super) population of students, rather than for a specific principal stratum. Unlike that case, however, these estimands do incorporate information about $Z_i$.

\subsection{Principal scores}
To proceed we require estimating different quantities conditional on specific values of $X$.
If $X$ is high-dimensional, this can be untenable as we would observe few units for any given value.
Borrowing from the propensity score literature, we can reduce the dimensionality of $X$ by calculating what is known as the \textit{principal score}~\citep{Hill:2002dt} for stratum $s$: 
$$\pi_{s\mid \mathbf{x}} \equiv \mathbb{P}\{S_i = \text{s} \mid \mathbf{X}_i = \mathbf{x}\}.$$
In our tutoring example, $\pi_{\text{h}\mid \mathbf{x}}$ is equivalent to $\mathbb{P}\{D_i(1) = \text{H} \mid \mathbf{X}_i = \mathbf{x}\}$, since principal stratum membership is entirely determined by $D_i(1)$ in the one-sided case. 
In addition, since there are only two strata in the one-sided case and since $\pi_{\text{h}\mid \mathbf{x}} = 1 - \pi_{\ell\mid \mathbf{x}}$, we can abuse notation somewhat and arbitrarily define the principal score for a given unit $i$ as $\pi_i = \pi_{h\mid \mathbf{X_i} = \mathbf{x}}$. 
As shown in Lemma~\ref{lem:principal_score}, below, principal scores shares two desirable properties with propensity scores. 
A proof of this lemma is in Appendix~\ref{sec:pscore_proofs}; see also~\citet{Jo:2009hx} and~\citet{Ding:2014wc}.

\begin{lem}[Properties of the Principal Score]\label{lem:principal_score}
The principal score, $\pi_i$, is a balancing score in the sense that
$S_i \ind \mathbf{X}_i \mid \pi_i.$
Furthermore, if either Strong or Weak Principal Ignorability holds given $\mathbf{X}_i$, that same assumption also holds given $\pi_i$.
\end{lem}

As a result, we can reduce the dimensionality of $\mathbf{X}$ to a scalar. 
Furthermore, in the one-sided case, we can directly estimate the principal score, since $\mathbb{P}\{S_i = \text{h} \mid Z_i = 1, \mathbf{X}_i = \mathbf{x}\} = \mathbb{P}\{S_i = \text{h} \mid Z_i = 0, \mathbf{X}_i = \mathbf{x}\} = \mathbb{P}\{S_i = \text{h} \mid \mathbf{X}_i = \mathbf{x}\}$ due to randomization. 
Therefore, asymptotically, we can obtain a non-parametric estimate of $\pi_i$ by estimating the proportion of $D_i(1) = \text{H}$ for each $\mathbf{X}_i = \mathbf{x}$. 
See \citet{Abadie:2003ve}.
Alternatively, following~\citet{Schochet:2007ky} and~\citet{Jo:2009hx}, this can also be done via modeling, such as with a logistic regression of $D^{\obs}$ on $\mathbf{X}$ among those students with $Z_i = 1$. 
Once we have a model, we can estimate $\widehat{\pi}_i$ for all students in the sample, including those with $Z_i = 0$.

Just as with the propensity score, a key concern is whether the principal score model has been correctly specified~\citep[see][]{imbens_rubin2014}. In the case of one-sided noncompliance, we can compare the covariate distribution for observed Low Takers and observed High Takers with $Z_i = 1$ and with similar values of $\widehat{\pi}$. Since the principal score is a balancing score, these distributions should be close, analogous to the propensity score setting. Poor balance is evidence of a mis-specified principal score model. As we discuss below, this is more complex with two-sided noncompliance. 

One interesting alternative to balance-checking presented in \citet{Ding:2014wc} is to directly compare the covariate distribution for those students assigned to the treatment group and \textit{observed} to be High Takers, and those students assigned to the control group \textit{predicted} to be  High Takers, with analogous comparisons for Low Takers. We can then compare means or other functions of these two groups to determine covariate balance. Poor balance between those assigned to treatment and those assigned to control within each principal stratum (either observed or predicted) is again evidence of a mis-specified principal score model.

\section{Estimation in the one-sided case\label{sec:one_sided_estimation}}
There are a variety of ways to estimate $\text{ITT}_{\text{h}}$ and  $\text{ITT}_{\text{$\ell$}}$ given either Weak or Strong Principal Ignorability. 
We focus on two methods proposed in the literature: the ``discrete subgroup" method of~\citet{Schochet:2007ky} and the ``weighting" method of~\citet{Jo:2009hx} and~\citet{Ding:2014wc}. 
Other methods exist that we do not address here, such as regression~\citep{Joffe:2007fk,Bein:2015ud} and matching~\citep{Hill:2002dt,Jo:2009hx}. For further discussion see~\citet{Porcher:2015bi}.
As these alternate methods are inherently driven by the critical principal ignorability assumptions, we anticipate that the intuition for the two methods we study should carry over.

Before turning to the general case, we first give some intuition for how the assumptions allow for estimation by examining the case of a single binary covariate.

\subsection{Estimation with a single, binary covariate}
Let $X$ be a single binary covariate, such as student's sex, $X_i \in \{m,f\}$. Let $p_{x} \equiv \mathbb{P}\{X_i = x\}$ be the proportion of students with $X_i = x$; $p_{x|s} = \mathbb{P}\{X_i = x \mid S_i = s\}$ be the proportion of students with $X_i = x$ among those with $S_i = s$; and let $\pi_{x} \equiv \mathbb{P}\{S_i = \text{h} \mid X_i = x\}$ be the probability that a randomly selected student with $X_i = x$ would take the High Dose.
The principal score for a student with $X_i = x$ is then $\pi_x$.
Since $X$ is binary, we can immediately estimate $\pi_x$ among those assigned to treatment via the observed proportion of $D_i^{\obs} = \text{H}$ for $Z_i = 1$ and $X_i = x$, with estimated quantities denoted $\widehat{\pi}_{x}$. 

We can also directly estimate four outcome means, $\overline{Y}_{1\mid m}$, $\overline{Y}_{1\mid f}$, $\overline{Y}_{0\mid m}$, and $\overline{Y}_{0\mid f}$, where $\overline{Y}_{z\mid x}$ is the average of those units with $Z_i = z$ and $X_i = x$.
For the treatment side, because we can identify the dose groups we can also estimate $\overline{Y}_{1\text{H}}$ and $\overline{Y}_{1\text{L}}$ as the average of those units who received treatment and took the high or low dose, respectively, as well as $\overline{Y}_{1\text{H}|f}, \overline{Y}_{1\text{H}|m}, \overline{Y}_{1\text{L}|m}$ and $\overline{Y}_{1\text{L}|m}$, which are the averages of the subgroups defined by receiving treatment ($Z_i = 1$), taking High or Low dose ($D_i = H$ or $L$), and having $X_i = f$ or $X_i = m$.

To estimate treatment effects for Low and High Takers, we need to estimate their average outcomes under both treatment and control.
We discuss how we do this by leveraging the ignorability assumptions next.

\subsubsection{Estimating average outcomes for $Z_i = 0$}
On the control side, we use $X$ to estimate two key quantities of interest, $\mu_{\text{h}0}$ and $\mu_{\text{$\ell$}0}$. 
Either Strong or Weak PI give $\mu_{\text{h}0 \mid f} = \mu_{\text{$\ell$}0 \mid f}$ and $\mu_{\text{h}0 \mid m} =  \mu_{\text{$\ell$}0 \mid m}$. 
Because of this $\overline{Y}_{0\mid f}$, the average outcome for those with $X_i=f$ in the control group, is an unbiased estimate of both $\mu_{\text{h}0 \mid f}$ and $\mu_{\text{$\ell$}0 \mid f}$. Same for $\overline{Y}_{0\mid m}$.

In addition, the overall mean of the High Takers in the control group, $\mu_{\text{h}0}$, can be expressed as a weighted average of the two subgroups defined by $X_i$:
$$\mu_{\text{h}0} = \frac{p_{f|h}}{p_{f|h} + p_{m|h}} \mu_{\text{h}0 \mid f} +\frac{p_{m|h}}{p_{f|h} + p_{m|h}} \mu_{\text{h}0 \mid m}.$$
The weights the relative size of these subgroups in the High Taker principal stratum.
Under principal ignorability, we can immediately estimate $\mu_{\text{h}0 \mid x}$ via $\overline{Y}_{0 \mid x}$. 
To estimate $p_{x|h}$ we apply Bayes Rule:
\begin{align*}
 p_{x|s} = \mathbb{P}\{X_i = x | S_i = s\} &= \frac{ \mathbb{P}\{ S_i = s | X_i = x \}~ \mathbb{P}\{ X_i = x \} }{ \mathbb{P}\{S_i = s\} } = \frac{ \pi_{s|x}~p_x}{ \pi_s }.
\end{align*}

The plug-in moment estimator for $\mu_{\text{h}0}$ is therefore
\begin{eqnarray*}
\widehat{\mu}_{\text{h}0} &=& \frac{\widehat{p}_{f|h}}{\widehat{p}_{f|h} + \widehat{p}_{m|h}} \widehat{\mu}_{\text{h}0 \mid f} + \frac{\widehat{p}_{m|h}}{\widehat{p}_{f|h} + \widehat{p}_{m|h}}\widehat{\mu}_{\text{h}0 \mid m}, \\
&=& \frac{\widehat{\pi}_f\widehat{p}_f}{\widehat{\pi}_f\widehat{p}_f + \widehat{\pi}_m\widehat{p}_m} \overline{Y}_{0 \mid f} + \frac{\widehat{\pi}_m\widehat{p}_m}{\widehat{\pi}_f\widehat{p}_f + \widehat{\pi}_m\widehat{p}_m} \overline{Y}_{0 \mid m},
\end{eqnarray*}
where the overall $\pi_h$ terms in the numerator and denominator cancel. In other words, we estimate $\mu_{\text{h}0}$ via the weighted average of the subgroup mean estimates, $\overline{Y}_{0 \mid f}$ and $\overline{Y}_{0 \mid m}$, with weights determined by the proportion of High Takers in each subgroup. We estimate $\mu_{\text{$\ell$}0}$ analogously.

The intuition behind the above is that ignorability states that two students with the same covariate $X$ will have the same control outcome on average, regardless of their observed take-up of dose. 
Thus, the overall mean for a subgroup of interest under the control condition is a weighted average of these predictions (the ``$\overline{Y}_{0\mid\text{x}}$''s), with weights determined by the distribution of $X$ in the subgroup of interest, which we observe on the treatment side (the ``$\widehat{p}_{x|s}$''s).

\subsubsection{Estimating average outcomes for $Z_i = 1$ and the final ITT estimates}
We next estimate the mean outcomes under treatment.
We can then subtract the control estimates from the previous section to obtain the ITT estimates.
We directly observe stratum membership for those individuals with $Z_i = 1$ because strata membership is fully determined by behavior under treatment (i.e., because of one-sided non-compliance). Under both Strong and Weak PI, we can therefore directly estimate $\mu_{\text{h}1}$ and $\mu_{\text{$\ell$}1}$ via the observed means for these two groups, $\overline{Y}_{1\text{H}}$ and $\overline{Y}_{1\text{L}}$, respectively. We now discuss possible estimators under Weak and Strong PI.

\paragraph{Weak PI.} Since Weak PI is only a statement about $Y_i(0)$ and not about $Y_i(1)$, we use the direct estimators $\overline{Y}_{1\text{H}}$ and $\overline{Y}_{1\text{L}}$ for $\mu_{\text{h}1}$ and $\mu_{\ell 1}$. This yields the following moment estimators for the quantities of interest:
\begin{align*}
\widehat{\text{ITT}}_{\text{h}} &= \widehat{\mu}_{\text{h}1} - \widehat{\mu}_{\text{h}0} =  \overline{Y}_{1\text{H}} - \left[\frac{\widehat{p}_{f | h}}{\widehat{p}_{m | h} + \widehat{p}_{f | h}} \overline{Y}_{0 \mid f} + \frac{\widehat{p}_{m | h}}{\widehat{p}_{m | h} + \widehat{p}_{f | h}} \overline{Y}_{0 \mid m}\right], \\
\widehat{\text{ITT}}_{\text{$\ell$}} &= \widehat{\mu}_{\text{$\ell$}1} - \widehat{\mu}_{\text{$\ell$}0} =  \overline{Y}_{1\text{L}} - \left[\frac{\widehat{p}_{f | \ell}}{\widehat{p}_{m | \ell} + \widehat{p}_{f | \ell}} \overline{Y}_{0 \mid f} + \frac{\widehat{p}_{m | \ell}}{\widehat{p}_{m | \ell} + \widehat{p}_{f | \ell}} \overline{Y}_{0 \mid m}\right].
\end{align*}

\paragraph{Strong PI.} First, we could use the Weak PI estimators.  
However, as all the necessary information about stratum membership is contained in $X$, there should be no difference between using the direct estimators $\overline{Y}_{1\text{H}}$ and $\overline{Y}_{1\text{L}}$ or the estimators that instead use information about $X_i$. 
In particular we could, just as with the control side, estimate $\mu_{\text{h}1}$ via the weighted average of subgroups:
$$\widehat{\mu}_{\text{h}1} = \frac{\widehat{p}_{f | h}}{\widehat{p}_{m | h} + \widehat{p}_{f | h}} \overline{Y}_{1 \mid f} + \frac{\widehat{p}_{m | h}}{\widehat{p}_{m | h} + \widehat{p}_{f | h}} \overline{Y}_{1 \mid m},$$
with an analogous estimator for $\mu_{\text{$\ell$}1}$. This yields:

\begin{align}
\nonumber \widehat{\text{ITT}}_{\text{h}} &=  \left[ \frac{\widehat{p}_{f | h}}{\widehat{p}_{m | h} + \widehat{p}_{f | h}} \overline{Y}_{1 \mid f} + \frac{\widehat{p}_{m | h}}{\widehat{p}_{m | h} + \widehat{p}_{f | h}} \overline{Y}_{1 \mid m}\right] - \left[\frac{\widehat{p}_{f | h}}{\widehat{p}_{m | h} + \widehat{p}_{f | h}} \overline{Y}_{0 \mid f} + \frac{\widehat{p}_{m | h}}{\widehat{p}_{m | h} + \widehat{p}_{f | h}} \overline{Y}_{0 \mid m}\right]  \\ 
\label{eq:ITThatc} &= \frac{\widehat{p}_{f | h}}{\widehat{p}_{m | h} + \widehat{p}_{f | h}} \left(\overline{Y}_{1 \mid f} - \overline{Y}_{0 \mid f}\right) + \frac{\widehat{p}_{m | h}}{\widehat{p}_{m | h} + \widehat{p}_{f | h}} \left(\overline{Y}_{1 \mid m} - \overline{Y}_{0 \mid m}\right)\\[2em] 
\nonumber \widehat{\text{ITT}}_{\ell} &=  \left[ \frac{\widehat{p}_{f | \ell}}{\widehat{p}_{m | \ell} + \widehat{p}_{f | \ell}} \overline{Y}_{1 \mid f} + \frac{\widehat{p}_{m | \ell}}{\widehat{p}_{m | \ell} + \widehat{p}_{f | \ell}} \overline{Y}_{1 \mid m}\right] - \left[\frac{\widehat{p}_{f | \ell}}{\widehat{p}_{m | \ell} + \widehat{p}_{f | \ell}} \overline{Y}_{0 \mid f} + \frac{\widehat{p}_{m | \ell}}{\widehat{p}_{m | \ell} + \widehat{p}_{f | \ell}} \overline{Y}_{0 \mid m}\right]  \\ 
\label{eq:ITThatn} &= \frac{\widehat{p}_{f | \ell}}{\widehat{p}_{m | \ell} + \widehat{p}_{f | \ell}} \left(\overline{Y}_{1 \mid f} - \overline{Y}_{0 \mid f}\right) + \frac{\widehat{p}_{m | \ell}}{\widehat{p}_{m | \ell} + \widehat{p}_{f | \ell}} \left(\overline{Y}_{1 \mid m} - \overline{Y}_{0 \mid m}\right) 
\end{align}
Unlike in the Weak PI case, these estimators are simple weighted averages of ITT estimates for subgroups defined by $X$, with weights $\widehat{p}_{x|s}$.
The only role stratum membership plays is in these weights.

Because we have two distinct estimators of the same thing, Strong PI yields a testable implication.
If the estimates obtained via the Weak and Strong PI assumptions are not equal beyond measurement error, the treatment side of the Strong PI assumption, $Y_i(1) \ind D_i(1) \mid \mathbf{X}_i$, must not hold. 
This test does not inform us, however, as to whether the control side of the assumption does or does not hold.

\subsection{Estimating impacts in general}
In our binary $X$ example, our estimators are weighted averages of subgroup means for subgroups defined by our covariate, with weights defined by the distribution of the covariate in our principal strata.
We obtain these weights as a function of the proportions of units in each principle strata for each value of $X$.
This approach can be readily extended to more general $X$.

Since we can directly observe the distribution of $X$ we can, in principle, immediately calculate subgroup means for any given $X$.
So this part is immediate.
The critical quantity, then, is what proportion of units are in a given principal strata for any given value of $X$.
We next discuss the details of this generalization, and then turn to two more practical approaches for estimating causal effects via principal scores.

\subsubsection{General setup}
The above readily extends to the general case with both discrete and continuous covariates.
First, we can directly estimate $\mu_{s1}$ by averaging our units with $Z_i=1$ and $D_i=s$.

We then estimate the stratum-specific mean $\mu_{s0}$ as a weighted average across an infinite number of subgroups defined by $X$ (i.e., an integral):
\[ \mu_{s0} = \EE{ \EE{ Y^{\obs}_i | S_i = s, Z_i = 0, X_i = x  } \mid X_i } = \EE{ \mu_{s0|x} | X_i } = \int_{x} \mu_{s0|x}~p_{x|s}~dF( x ), \]
where $p_{x|s} = \mathbb{P}[X_i = x \mid S_i = s]$ and $\mu_{s0|x} = \EE{ Y_i(0) | S_i = s, X_i = x}$. This is simply the Law of Iterated Expectations conditional on strata membership, where randomization allows us to drop the conditioning on $Z$. As in the binary case above, we can use Bayes Rule to replace $p_{x|s}$ by $\pi_{s|x}p_x/\pi_s$ giving
\begin{equation}
 \mu_{s0} = \int_{x} \mu_{s0|x} p_{x|s} dF( x ) =  \int_x \mu_{s0|x} \cdot \pi_{s|x} \cdot \frac{ 1 }{ \pi_s }  dF( x ),
\end{equation}
where $dF(x)$ is the distribution of $X$ in the population (i.e., $p_x$). 

We next need to estimate the components of this integral.  
There are different methods for doing this.
If we estimate the distribution $dF(x)$ with the empirical distribution and use an estimated prognostic score model for the $\pi_{s|x}$ this integral can be estimated as a summation over all $N$ units, with individual weights $\widehat{\pi}_i = \widehat{\pi}_{s|X_i}$:
\begin{equation}
 \widehat{\mu}_{s0} = \frac{\sum_i \widehat{\pi}_i \cdot \widehat{\mu}_{s0|X_i}}{\sum_i \widehat{\pi}_i},
\end{equation}
using a natural estimate of $\widehat{\pi}_s = \sum \widehat{\pi}_i / N$.

The practical question is then how best to estimate $\widehat{\mu}_{s0|X_i}$. In the case of discrete $X$, as in the previous section, the subgroup mean, $\overline{Y}_{0|x}$, is a natural estimator. More broadly, we could use nonparametric regression to estimate this quantity. As we discuss next, however, a straightforward approach is simply to use the observed outcome here.

\subsubsection{Weighting method}

We now turn to approaches that re-weight the observed outcomes directly. Here we estimate $\mu_{\text{h}0}$ only using individuals assigned to control:
$$
\widehat{\mu}_{\text{h}0} = \frac{1}{\sum_{i: Z_i = 0}  \widehat{\pi}_i} \sum_{i: Z_i = 0}  \widehat{\pi}_i Y^{\obs}_i .
$$
This works because $Y_i^\obs$ estimates $\mu_{s0\mid X_i}$ for units with $Z_i=0$.
Take the average of units on the treatment side with $D_i=H$ to estimate $\mu_{h1}$ and we have an overall estimate of $\widehat{\text{ITT}}_{\text{h}}$ under Weak PI of
$$\widehat{\text{ITT}}_{\text{h}} =\left[\frac{1}{N_{1\text{H}}} \sum_{i: Z_i = 1, D_i^{\obs} = \text{H}} Y_i^{\obs}\right] - \left[\frac{1}{\sum_{i: Z_i = 0}  \widehat{\pi}_i} \sum_{i: Z_i = 0}  \widehat{\pi}_i Y^{\obs}_i \right].$$
$\widehat{\text{ITT}}_{\text{h}}$ is simply a weighted difference in means estimator with weights
	$$w_i = \begin{cases} 
	1 & \text{if } Z_i = 1 \text{ and } D_i^{\text{obs}} = \text{H} \\
	0 & \text{if } Z_i = 1 \text{ and } D_i^{\text{obs}} = \text{L} \\
	\widehat{\pi}_i & \text{if } Z_i = 0
	\end{cases}\qquad.$$
This is the weighting estimator proposed by~\citet{Jo:2009hx}. 

We can easily extend this to the case of Strong PI by using a similar expression to the control weighted average, obtaining
\begin{align*}
\widehat{\text{ITT}}_{\text{h}} 
&= \left[\frac{1}{\sum_{i: Z_i = 1}  \widehat{\pi}_i} \sum_{i: Z_i = 1}  \widehat{\pi}_i Y^{\obs}_i \right] - \left[\frac{1}{\sum_{i: Z_i = 0}  \widehat{\pi}_i} \sum_{i: Z_i = 0}  \widehat{\pi}_i Y^{\obs}_i \right] ,
\end{align*}
which is again a weighted difference in means estimator with weight $w_i \equiv \widehat{\pi}_i$ for all students, regardless of treatment assignment.


\subsubsection{Discrete subgroup method}
\citet{Schochet:2007ky} propose a straightforward approach for estimating $\text{ITT}_\text{h}$ and $\text{ITT}_\text{$\ell$}$. First, let $\bar{\pi} \equiv \mathbb{P}\{S_i = \text{h}\}$ be the overall proportion of High Takers in the population, with corresponding moment estimate $\widehat{\bar{\pi}}$. Next, define $\widehat{H}_i = \mathbb{I}\{ \widehat{\pi}_i \geq \widehat{\bar{\pi}}\}$, the indicator for whether student $i$ is predicted to be a High Taker based on being above a given threshold. 
Finally estimate $\text{ITT}_\text{h}$ as the estimated ITT impact for those students with $\widehat{H}_i = 1$ and estimate $\text{ITT}_\text{$\ell$}$ as the estimated ITT impact for those students with $\widehat{H}_i = 0$.

The intuition here is our predictive model does not depend on outcomes or treatment assignment.
The identified subgroups, which we might call ``Likely High Dose'' and ``Likely Low Dose,'' are therefore pre-treatment subgroups and can be described and explored just as any other pre-treatment subgroup.
Having such easily interpretable groups, and being able to leverage straightforward estimation procedures on them, is appealing.

To illustrate this approach, we turn back to the simple case with binary $X$.
First, without loss of generality, assume that students with $X_i = f$ are more likely to take a High dose than those with $X_i = m$, i.e., $\pi_{f} > \pi_{m}$.
Our original estimate was made by weighting the covariate defined subgroups as in Equation~\ref{eq:ITThatc}.
The discrete subgroup method, by contrast, is to only use the first term of this equation to estimate $\text{ITT}_\text{h}$:
$$\widehat{\text{ITT}}_{\text{h}}^{\text{Sub}} =\overline{Y}_{1 \mid f} - \overline{Y}_{0 \mid f}.$$
That is, we use the estimated ITT among women as a proxy for $\text{ITT}_{\text{h}}$. This only matches the plug-in estimator if $X_i$ is perfectly predictive of $S_i$ (i.e., $\pi_1 = 1$), or if there is no impact variation across principal strata (i.e., $\text{ITT} = \text{ITT}_{\text{h}} = \text{ITT}_{\text{$\ell$}}$), neither of which is an interesting case. 
While it might be possible to motivate this estimator with a different set of assumptions, these are not immediately apparent. 
These quantity are, however, valid estimates for alternate estimands: the average effects for groups defined by predicted membership.
For example, the estimate for $ITT_h$ is an estimate for the average impact of those predicted to be likely High Takers.

\subsection{Simulation Study}
We now present the results of a small simulation study that assess the finite sample properties of three approaches for estimating $\text{ITT}_{\text{h}}$:
\begin{itemize}
	\item \textbf{Discrete subgroup method.} This is the method proposed by~\citet{Schochet:2007ky}.
	\item \textbf{Weighting under Strong PI.} This is a modified version of the method first proposed by~\citet{Jo:2009hx}, with weight $w_i = \widehat{\pi}_i$ for all students.
	\item \textbf{Weighting under Weak PI.} This is the method first proposed by~\citet{Jo:2009hx}, with weight $w_i = \widehat{\pi}_i$ for all students assigned to control, weight $w_i = 1$ for all students assigned to treatment observed to be High Takers, and weight $w_i = 0$ for all students assigned to treatment observed to be Low Takers.
\end{itemize}

Mirroring the simulation study in~\citet{Stuart:2011gd}, we generate strata membership, $H_i$, and outcome data, $Y_i(1)$ and $Y_i(0)$, from the following model:\footnote{There are obviously many ways to parameterize such a simulation study.~\citet{Schochet:2007ky}, for example, generate data from a standard selection model in which they vary the correlation of the error terms between the selection model and the outcome equation.}
\begin{eqnarray*}
x_i &\sim & N(0,1) \\ 
\text{logit}^{-1}(\pi_i) &=& \eta_0 + \eta_1 x_i \\
H_i &\sim & Bern( \pi_i ) \\
Y_i(0) &=& \alpha + \beta_0 x_i + \gamma_0 H_i + \delta_0 H_i x_i + \varepsilon_{y,i} \\
\tau_i &=& \tau + \beta_1 x_i + \gamma_1 H_i + \delta_1 H_i x_i + \varepsilon_{\tau, i}
\end{eqnarray*}
with $Y_i(1) = Y_i(0) + \tau_i$.
In this model $H_i$ is an indicator for whether student $i$ is a High Taker, and is generated via a logistic function.
The residual noise terms are distributed as $\varepsilon_{y,i} \sim N(0, \sigma^2_y)$, and $\varepsilon_{\tau,i} \sim N(0, \sigma^2_\tau)$. The key parameters for Principal Ignorability are $\gamma$ and $\delta$. 
In all simulations, our covariate does predict strata membership.  
The question is how it is connected to outcomes.
Under Strong Principal Ignorability, $\gamma_0 = \gamma_1 = 0$, and $\delta_0 = \delta_1 = 0$, so $H_i$ does not impact outcomes at all, only $x$ does; under Weak Principal Ignorability, $\gamma_0 = 0$ and $\delta_0 = 0 $, but $\gamma_1$ and $\delta_1$ are unconstrained. 

In these simulations we, to avoid the increased complexity of interaction terms, always set $\delta_\ell = 0$ and only manipulate violations of our assumption via the $\gamma_\ell$.
We then set the following parameter values to be common across simulations: $\alpha = 0$, $\beta_0 = 0.5$, $\tau = 0.5$, and $\sigma_\tau = 0.1$.
We set the residual variance of $\sigma_y = 1$ as well.
Finally, we explore sensitivity to $\beta_1$, letting it range from 0 to 0.25 giving a range of no systematic treatment variation dependent on $X$ to substantial variation.
Our final data generation model is then
\begin{eqnarray*}
Y_i(0) &=& 0.5 x_i + \gamma_0 H_i + \varepsilon_{y,i} \\
\tau_i &=& 0.5 + \beta_1 x_i + \gamma_1 H_i + \varepsilon_{\tau, i} .
\end{eqnarray*}

For strata membership, we set $\eta_0 = 0$, so that the overall proportion of High Takers in each generated data set is 50\% in expectation. 
We conducted simulations for different $\eta_1$, but the result were largely insensitive to this parameter. 
For ease of presentation, we therefore only show results with $\eta_1 = 1$.

For each simulation run, we generate 1,000 data sets from the above Data Generating Process, each with $N =$ 2,000 and $p = 0.5$ randomly assigned to treatment.

\paragraph{Simulations under Strong Principal Ignorability.} Table~\ref{tbl:sim_w_PI} shows the results of the simulation study when Strong Principal Ignorability holds ($\gamma_0 = \gamma_1 = 0$). The first row shows results for the case with no impact variation across either $X$ or across principal strata.
In this case $\text{ITT}_\text{h} = \text{ITT}_\text{$\ell$}$ and, unsurprisingly, all three methods are unbiased and have good coverage. 
The second and third rows show the case in which Strong Principal Ignorability still holds, but in which there is also impact variation across $X$, that is, $\beta_1 \neq 0$, which makes $\text{ITT}_\text{h} \neq \text{ITT}_\text{$\ell$}$.
In these cases, the weighting methods continue to perform well. However, the discrete subgroup method is biased and has poor coverage.

\begin{table}
	\begin{center}
	\begin{threeparttable}
	\centering
					\caption{\label{tbl:sim_w_PI} Simulation studies under Strong Principal Ignorability, with varying $\beta_1$.}
	\begin{tabular}{ccc c ccc c ccc}
	\toprule
	& & & & \multicolumn{3}{c}{\textbf{Bias}} & &  \multicolumn{3}{c}{\textbf{95\% Coverage}} \\
\cline{5-7} \cline{9-11}
$\beta_1$ & $\gamma_0$ & $\gamma_1$ &  & \textit{Sub.} & \textit{Wt.} & \textit{Wk. Wt.} & & \textit{Sub.} & \textit{Wt.} & \textit{Wk. Wt.} \\
\midrule
0 & 0 & 0 & & 0.00 & 0.00 & 0.00 &  & 0.95 & 0.93 & 0.94 \\
0.1 & 0 & 0 & & 0.04 & 0.00 & 0.00 &  & 0.91 & 0.94 & 0.94 \\
0.25 & 0 & 0 & & 0.10 & 0.00 & 0.00 &  & 0.68 & 0.93 & 0.93 \\
\bottomrule
\end{tabular}
\begin{tablenotes}
\item \scriptsize \textit{Note:} $\eta_1 = 1$; results for $\text{ITT}_{\text{h}}$; ``Sub" is the discrete subgroup method; ``Wt." is the weighting method under Strong PI; and ``Wk Wt." is the weighting method under Weak PI.
\end{tablenotes}
\end{threeparttable}
\end{center}
\end{table}

\paragraph{Simulations without Principal Ignorability.} Table~\ref{tbl:sim_wo_PI} shows the results of the simulation study when Principal Ignorability does not hold (for simplicity, $\beta_1 = 0$ throughout). As reference, the first row presents results under Strong PI, which again shows that all three methods are unbiased and have good coverage in this simple case.
This is same row (up to simulation error) as row 1 of Table~\ref{tbl:sim_w_PI}.
The rest of the first bank shows results for the case in which Weak PI holds but Strong PI does not, i.e., $\gamma_0 = 0$, $\gamma_1 \neq 0$.
When $\gamma_1 \neq 0$, knowledge of $X_i$ does not fully explain the treatment outcome, which causes this violation.
Unsurprisingly, both the discrete subgroup method and the weighting method under Strong PI are biased and have poor coverage.
The weighting method under Weak PI, however, performs well, as it estimates mean treatment outcomes directly.

The next two banks show results for settings in which neither Weak PI nor Strong PI holds.
Because $\gamma_0 \neq 0$, knowledge of $X_i$ does not fully explain the control outcome, which is the core assumption used in all our estimators.
In general, all three methods are biased and have poor coverage in these scenarios. 

In the degenerate case when $\gamma_1 = 0$ there is no impact variation across principal strata---$\text{ITT} = \text{ITT}_{\text{h}} = \text{ITT}_{\text{$\ell$}}$---although the component means differ. 
In this case both the subgroup method and weighting under Strong PI perform very well, even while the weighting under Weak PI does not.
This occurs because, even though we are applying the same (wrong) weights to units assigned to treatment and control, we will get reasonable estimates of the treatment effects because the treatment effect of any arbitrary subgroup will be the same as any other when there is no actual treatment effect variation.
Because the Weak PI case does not use weights on the treatment side, it is in effect comparing a different subgroup in control to the correct one in treatment.


\begin{table}
	\begin{center}
	\begin{threeparttable}
	\centering
		\caption{\label{tbl:sim_wo_PI} Simulation studies without Principal Ignorability.}
	\begin{tabular}{cc c ccc c ccc}
	\toprule
	& & & \multicolumn{3}{c}{\textbf{Bias}} & &  \multicolumn{3}{c}{\textbf{95\% Coverage}} \\
\cline{4-6} \cline{8-10}
$\gamma_0$ & $\gamma_1$ &  & \textit{Sub.} & \textit{Wt.} & \textit{Wk. Wt.} & & \textit{Sub.} & \textit{Wt.} & \textit{Wk. Wt.} \\
\midrule
0 & 0 & & 0.00 & 0.00 & 0.00 &  & 0.95 & 0.95 & 0.94 \\
0 & 0.2 & & -0.10 & -0.10 & 0.00 &  & 0.68 & 0.48 & 0.95 \\
0 & 0.5 & & -0.24 & -0.25 & 0.00 &  & 0.06 & 0.00 & 0.95 \\[1em]
0.2 & 0.0 & & 0.00 & 0.00 & 0.10 &  & 0.95 & 0.94 & 0.61 \\
0.2 & 0.2 & & -0.10 & -0.10 & 0.10 &  & 0.71 & 0.50 & 0.59 \\
0.2 & 0.5 & & -0.24 & -0.25 & 0.10 & & 0.04 & 0.00 & 0.62 \\[1em]
0.5 & 0.0 & & 0.00 & 0.00 & 0.25 &  & 0.95 & 0.96 & 0.01 \\
0.5 & 0.2 & & -0.09 & -0.10 & 0.25 &  & 0.71 & 0.53 & 0.01 \\
0.5 & 0.5 & & -0.24 & -0.25 & 0.25 & & 0.09 & 0.00 & 0.01 \\
\bottomrule
\end{tabular}
\begin{tablenotes}
\item \scriptsize \textit{Note:} $\eta_1 = 1$; $\beta_1 = 0$; results for $\text{ITT}_{\text{h}}$; ``Sub" is the discrete subgroup method; ``Wt." is the weighting method under Strong PI; and ``Wk Wt." is the weighting method under Weak PI.
\end{tablenotes}
\end{threeparttable}
\end{center}

\end{table}


\section{Principal Ignorability in the two-sided case\label{sec:two_sided}}
We next discuss extensions of the above assumptions and methods to the two-sided case, that is, when $D_i(0)$ is not constant for all units. 
We also demonstrate how to combine the Weak Principal Ignorability assumption with an exclusion restriction for one of the principal strata of interest. 
We illustrate this form of two-sided noncompliance with an extension of our binary covariate example to show how these formula look in practice.
Finally, we apply that approach to an applied example.

\subsection{Setup}
We use the Head Start Impact Study~\citep{puma2010hsis} as our running example for two-sided noncompliance, discussed in more detail below. 
Let $Z_i$ denote whether child $i$ is randomly offered the opportunity to enroll in Head Start; $Y_i$ denote child $i$'s outcome of interest, which we will set as the Peabody Picture Vocabulary Test (PPVT) score; and $\mathbf{x}_i$ denote a vector of pre-treatment covariates, including pre-test score. 
Let $D_i \in \{0, 1\}$ be an indicator for whether child $i$ enrolls in Head Start. 
The substantive question of interest is the effect of enrolling in Head Start. 
Finally, we invoke the \textit{monotonicity} or ``no defiers'' assumption, which assumes that the offer of enrollment in Head Start did not induce any children to do the opposite. 
This yields three possible principal strata:
$$ S_i = \begin{cases}
	\text{Always Taker (a)}  & \text{if } D_i(0) = 1 \text{ and } D_i(1) = 1 \\
	\text{Complier (c)}  & \text{if } D_i(0) = 0 \text{ and } D_i(1) = 1 \\
	\text{Never Taker (n)}  & \text{if } D_i(0) = 0 \text{ and } D_i(1) = 0,
	\end{cases}$$
where $S_i = \text{a}$, $S_i = \text{h}$, and $S_i = \text{n}$ indicate that individual $i$ is an Always Taker, Complier, and Never Taker, respectively.
Comparing this with our tutoring example, we see the language of complier vs. never taker more explicitly here: students are offered treatment or not, and they end up taking treatment or not. 
Unlike traditional non-compliance, however, we leave room for the possibility of a treatment effect of being offered treatment in addition to taking treatment.
We codify this possibility in ignorability assumptions as before.  
We discuss this next.

Table~\ref{tbl:twosided_example} shows the relationship between observed groups and principal strata in this example. 
Analogous to the one-sided case in Table~\ref{tbl:onesided_example}, we can immediately estimate the overall proportion of each principal stratum: $\pi_\text{a} = \mathbb{P}\{D_i(0) = 1 \mid Z_i = 0\}$, $\pi_\text{n} = \mathbb{P}\{D_i(1) = 0 \mid Z_i = 1\}$, and $\pi_\text{c} = 1 - \pi_\text{a} - \pi_\text{a}$. 
We can also immediately estimate $\mu_{\text{a}0}$ via the observed average outcomes for $Z_i = 0$, $D_i^{\obs} = 1$, denoted $\overline{Y}_{01}$, and $\mu_{\text{n}1}$ via the observed outcomes for $Z_i = 1$, $D_i^{\obs} = 0$, denoted $\overline{Y}_{10}$. 
However, we now have two mixtures to disentangle: the mixture of compliers and always-takers in the treatment group, and the mixture of compliers and never-takers in the control group.
We can observe the overall mean of these mixtures, but not the stratum-specific means.

The primary estimand of interest is $\text{ITT}_{\text{c}}$, the effect of enrolling in Head Start for those children who would enroll if offered the opportunity to do so and would not enroll if not offered. 
We are also interested in $\text{ITT}_\text{a}$, the effect of the offer of enrollment on children who would enroll in Head Start regardless of treatment assignment. 
As above, we will explore assumptions on the conditional outcome distributions that will allow us to estimate the causal effects of interest. 

	\begin{table}[btp]
	\centering
	\begin{threeparttable}
	\centering
			\caption{\label{tbl:twosided_example} Possible principal strata in two-sided example, under monotonicity.}
	\begin{tabular}{ccl}
			\toprule
			$Z_i$ & $D_i^{\text{ obs}}$ &  \multicolumn{1}{c}{\textbf{Possible principal strata}}\\
			\midrule
			$1$ & $1$ & Complier (treatment), Always Taker (treatment) \\
			$1$ & $0$ & Never Taker (treatment) \\
			$0$ & $1$ & Always Taker (control)\\
			$0$ & $0$ & Complier (control); Never Taker (control)\\
			\bottomrule
		\end{tabular}
	\end{threeparttable}
	\end{table}

\subsection{Principal Ignorability in the two-sided case}
The prior assumptions extend naturally to the two-sided case. 
As above, we can either assume Strong or Weak Principal Ignorability. 
In addition, we can combine Weak PI with an exclusion restriction.

\paragraph{Strong Principal Ignorability.} The Strong Principal Ignorability assumption is quite similar to the one-sided case:
\begin{align*}
Y_i(1) &\ind (D_i(1), D_i(0)) \mid \mathbf{X}_i, \\
Y_i(0) &\ind (D_i(1), D_i(0)) \mid \mathbf{X}_i.
\end{align*}
As in the one-sided case, this states that, given covariates, stratum membership is as good as randomly assigned.
Written in terms of mean-independence, we have:
\begin{align*}
\mathbb{E}[Y_i(1) \mid \mathbf{X}_i = \mathbf{x}, S_i = \text{a}] &= \mathbb{E}[Y_i(1) \mid \mathbf{X}_i = \mathbf{x}, S_i = \text{n}] = \mathbb{E}[Y_i(1) \mid \mathbf{X}_i = \mathbf{x}, S_i = \text{c}],\\
\mathbb{E}[Y_i(0) \mid \mathbf{X}_i = \mathbf{x}, S_i = \text{a}] &= \mathbb{E}[Y_i(0) \mid \mathbf{X}_i = \mathbf{x}, S_i = \text{n}] = \mathbb{E}[Y_i(0) \mid \mathbf{X}_i = \mathbf{x}, S_i = \text{c}].\\
\end{align*}
Or, more compactly:
\begin{align*}
	\mu_{\text{c}1 \mid \mathbf{x}} &= \mu_{\text{n}1 \mid \mathbf{x}} = \mu_{\text{a}1 \mid \mathbf{x}} = \mu_{1 \mid \mathbf{x}},\\ 
	\mu_{\text{c}0 \mid \mathbf{x}} &= \mu_{\text{n}0 \mid \mathbf{x}} = \mu_{\text{a}0 \mid \mathbf{x}} = \mu_{0\mid \mathbf{x}}.
\end{align*}
The observed means, $\overline{Y}_{0 \mid \mathbf{x}}$ and $\overline{Y}_{1 \mid \mathbf{x}}$, are the means by treatment \textit{assigned} (i.e., $Z$) and do not incorporate any information about treatment \textit{received} (i.e., $D$). 
In other words, under Strong PI, the average outcome depends only on $\mathbf{X}$ and $Z$---and not on $D$.
In particular, for our case this means that, given covariates, a students outcome does not depend on whether they attended a head-start center or not.
Clearly, this is a strong assumption.

\paragraph{Weak Principal Ignorability.} The Weak Principal Ignorability assumption differs from the one-sided case.
In particular, in this setting we directly observe Always Takers assigned to control and Never Takers assigned to treatment. 
This yields a pair of Weak Principal Ignorability assumptions:
\begin{align*}
Y_i(1) &\ind D_i(0) \mid \mathbf{X}_i, D_i(1) = 1, \\
Y_i(0) &\ind D_i(1) \mid \mathbf{X}_i, D_i(0) = 0.
\end{align*}
Re-written in terms of mean-independence:
\begin{align*}
	\mu_{\text{c}1 \mid \mathbf{x}} &= \mu_{\text{a}1 \mid \mathbf{x}} = \mu_{11 \mid \mathbf{x}},\\ 
	\mu_{\text{c}0 \mid \mathbf{x}} &= \mu_{\text{n}0 \mid \mathbf{x}} = \mu_{00 \mid \mathbf{x}},
\end{align*}
where $\mu_{zd|\mathbf{x}} = \mathbb{E}\{ Y_i^{\obs} | \mathbf{X}_i = \mathbf{x}, Z_i = z, D_i^{\obs} = d\}$. In words, given $\mathbf{X}$, Always Takers and Compliers assigned to treatment have the same average outcome; and, given $\mathbf{X}$, Never Takers and Compliers assigned to control have the same average outcome. 

These equalities are for units within observationally indistinguishable groups. 
Always Takers and Compliers assigned to treatment are all enrolled in Head Start. 
Never Takers and Compliers assigned to control are not enrolled in Head Start. 
This pair of assumptions states that, given $\mathbf{X}$, their counterfactual care setting is unrelated to their outcome in the observed care setting.
In other words, for a student known to be an Always Taker or Complier, our prediction of that student's outcome under the offer of treatment would not change with additional knowledge of which type of student they happen to be.

\paragraph{Exclusion Restriction and Weak Principal Ignorability.} An interesting extension is to replace one of the two conditional independence assumptions in Weak PI with an exclusion restriction. 
For example, in the Head Start scenario we can assume that there is no effect of the offer of enrollment on those children who would never enroll in Head Start regardless of treatment assignment; that is, we invoke the exclusion restriction for Never Takers. 
This yields:
 \begin{align}
Y_i(1) &\ind D_i(0) \mid \mathbf{X}_i, D_i(1) = 1, \label{eq:two_sided_weak_assumption}\\
Y_i(0) &= Y_i(1) \mbox{ for } S_i = \text{n}, \nonumber
\end{align}
or, in terms of mean independence,
$$\mu_{\text{c}1 \mid \mathbf{x}} = \mu_{\text{a}1 \mid \mathbf{x}} = \mu_{11 \mid \mathbf{x}},$$ 
$$\mu_{\text{n}0} = \mu_{\text{n}1}.$$
The first line is the PI for Always Takers and Compliers assigned to treatment. 
We have simply replaced the second Weak PI assumption with an exclusion restriction.


\subsection{Estimation with a binary covariate}
We illustrate the key ideas for the case with an exclusion restriction for the Never Takers and weak PI for the Always Takers (see Equation~\ref{eq:two_sided_weak_assumption}) with a single, binary $X \in \{m,f\}$, with $\mathbb{P}\{X_i = \text{m}\} = 1/2$. We focus on estimating the impact of randomization among Compliers, $\text{ITT}_{\text{c}}$, with similar results for the impact on Always Takers, $\text{ITT}_{\text{a}}$. 

First, estimating $\mu_{c0}$ is straightforward and does not involve covariates. Due to the exclusion restriction for Never Takers, we have that:
$$\mu_{10} = \mu_{n0} = \mu_{n1}.$$
Because the overall mean $\mu_{00}$ is a weighted average,
$$\mu_{00} = \mu_{\text{c}0} \frac{\pi_\text{c}}{\pi_\text{c} + \pi_{\text{n}}} + \mu_{\text{n}0} \frac{\pi_\text{n}}{\pi_\text{c} + \pi_{\text{n}}},$$
we can immediately estimate $\mu_{\text{c}0}$ via
$$\widehat{\mu}_{\text{c}0} = \overline{Y}_{00} \frac{\widehat{\pi}_\text{c} + \widehat{\pi}_{\text{n}} }{\widehat{\pi}_\text{c}} - \overline{Y}_{10} \frac{\widehat{\pi}_{\text{n}}}{\widehat{\pi}_\text{c} }.$$

To estimate $\mu_{\text{c}1}$, we need leverage the weak Principal Ignorability assumption. 
Under weak PI, $\mu_{\text{c}1 \mid x} = \mu_{\text{a}1 \mid x} = \mu_{11 \mid x}.$ Therefore, we can estimate the overall stratum  mean, $\mu_{\text{c}1}$, via the weighted average of $\mu_{\text{c}1 \mid m}$ and  $\mu_{\text{c}1 \mid f}$:

$$\widehat{\mu}_{\text{c}1} = \left( \frac{\widehat{p}_{m | \text{c}}  }{\widehat{p}_{m | \text{c}} + \widehat{p}_{f | \text{c}}} \right) \overline{Y}_{11 \mid m} + \left(\frac{\widehat{p}_{f | \text{c}}  }{\widehat{p}_{m | \text{c}} + \widehat{p}_{f | \text{c}}} \right)\overline{Y}_{11 \mid f},$$
where $p_{m | \text{c}} = \mathbb{P}\{X_i = m \mid S_i = \text{c}\}$.
Finally, we combine to obtain the overall estimator of $\text{ITT}_{\text{c}}$:

$$\widehat{\text{ITT}}_{\text{c}} = \left[\left( \frac{\widehat{p}_{m | \text{c}}  }{\widehat{p}_{m | \text{c}} + \widehat{p}_{f | \text{c}}} \right) \overline{Y}_{11 \mid m} + \left(\frac{\widehat{p}_{f | \text{c}}  }{\widehat{p}_{m | \text{c}} + \widehat{p}_{f | \text{c}}} \right)\overline{Y}_{11 \mid f}\right] - \left[ \overline{Y}_{00} \frac{\widehat{\pi}_\text{c} + \widehat{\pi}_{\text{n}} }{\widehat{\pi}_\text{c}} - \bar{Y}_{10} \frac{\widehat{\pi}_{\text{n}}}{\widehat{\pi}_\text{c} } \right].$$

\subsection{Estimation using principal scores}
\label{sec:est_score_two_sided}

Unsurprisingly, estimation is more complicated in the two-sided case. We first discuss estimating the principal score in this setting and then turn to estimating causal effects under various assumptions.

\subsubsection{Estimating the principal score in the two-sided case}
In the case of one-sided noncompliance, we directly observe stratum membership among those individuals assigned to treatment. In the two-sided case, however, we need an indirect approach as we never observe Compliers directly. We describe two broad estimation methods: marginal and joint estimation. We then briefly discuss model checking in this setting.

\paragraph{Marginal principal score estimation.} This approach takes advantage of the useful fact that we can directly observe Never Takers assigned to treatment and Always Takers assigned to control. In particular, we can directly estimate $\pi_{a \mid \mathbf{x}} \equiv \mathbb{P}\{S_i = \text{a} \mid \mathbf{X}_i = \mathbf{x}\}$ via the predicted probability from a logistic regression of $D$ on $\mathbf{X}$ in the control group. Similarly, we can estimate 
$\pi_{n \mid \mathbf{x}} \equiv \mathbb{P}\{S_i = \text{n} \mid \mathbf{X}_i = \mathbf{x}\}$ via 1 minus the predicted probability from a logistic regression of $D$ on $\mathbf{X}$ in the treatment group. Then, by construction, $\widehat{\pi}_{c \mid \mathbf{x}} = 1 - \widehat{\pi}_{n \mid \mathbf{x}} - \widehat{\pi}_{a \mid \mathbf{x}}$. Of course, we could replace logistic regression with nonparametric regression or similar estimation approaches.

\paragraph{Joint principal score estimation.} An obvious concern is that separately estimating $\widehat{\pi}_{a \mid \mathbf{x}}$ and $\widehat{\pi}_{n \mid \mathbf{x}}$ could lead to estimates for $\widehat{\pi}_{c \mid \mathbf{x}}$ that are outside $[0,1]$.  We can impose this constraint by jointly estimating the principal score models. For details on data augmentation, we refer interested readers to \citet{Ding:2014wc}, who illustrate a straightforward data augmentation procedure for estimation in this context. See also,~\citet{Ibrahim:1990im, Zhang:2009ku, Aronow:2013et,Hsu:2014un}. The key idea is to alternate between two steps. Given an initial vector of compliance types, repeat the following steps until convergence:
\begin{itemize}
	\item \textbf{Estimate the principal score.} Given the vector of compliance types, estimate the principal score via multinomial logistic regression, ignoring treatment assignment.
	\item \textbf{Impute compliance type.} Given the principal score model, impute compliance types for all individuals with unknown type. For EM, this is via maximization. For MCMC, this is via missing data imputation.
\end{itemize}
This procedure is essentially the model-based framework outlined in~\citet{Imbens:1997bayes} and~\citet{Hirano:2000vg}, but without including outcome information. 

\paragraph{Model checking.} Regardless of estimation strategy, we must assess whether we have a sensible model fit in practice. Following~\citet{Ding:2014wc}, we can compare the covariate distributions for observed and predicted Always Takers and Never Takers. For the Compliers, we can compare only compare the predicted distributions under treatment and control. 

Unfortunately, it is less useful to leverage the fact that the principal score is a balancing score in this setting. First, the principal score is now a vector, so we must consider units that are similar across two dimensions rather than one. Second, we can no longer observe Compliers directly, instead observing mixtures of Compliers and Never Takers or of Compliers and Always Takers. Thus, we prefer the balance checks in~\citet{Ding:2014wc} for two-sided noncompliance.


\subsubsection{Estimating effects via the principal score in the two-sided case}
We consider estimation under three types of assumption: (1) Strong PI, (2) Weak PI, and (3) Weak PI for Compliers and Always Takers and the exclusion restriction for Never Takers. 

\paragraph{Strong PI.} Since observed treatment received, $D$, is irrelevant under Strong PI, estimation is straightforward. To estimate the impact for stratum $s$, calculate the weighted average of units under treatment and control with weights $w_i = \widehat{\pi}_{i:s}$, just as in the one-sided case.

\paragraph{Weak PI.} We now condition on observed treatment received, $D$. For clarity, we write weights separately for each of the three principal strata.

$$w_{i:\text{a}} = \begin{cases} 
	\frac{\widehat{\pi}_{i:\text{a}}}{\widehat{\pi}_{i:\text{c}} + \widehat{\pi}_{i:\text{a}}}  & \text{if } Z_i = 1 \text{ and } D_i^{\text{obs}} = 1 \\
	0 & \text{if } Z_i = 1 \text{ and } D_i^{\text{obs}} = 0 \\
	1 & \text{if } Z_i = 0 \text{ and } D_i^{\text{obs}} = 1 \\
	0  & \text{if } Z_i = 0 \text{ and } D_i^{\text{obs}} = 0,
	\end{cases}$$

$$w_{i:\text{n}} = \begin{cases} 
	0  & \text{if } Z_i = 1 \text{ and } D_i^{\text{obs}} = 1 \\
	1 & \text{if } Z_i = 1 \text{ and } D_i^{\text{obs}} = 0 \\
	0 & \text{if } Z_i = 0 \text{ and } D_i^{\text{obs}} = 1 \\
	\frac{\widehat{\pi}_{i:\text{n}}}{\widehat{\pi}_{i:\text{c}} + \widehat{\pi}_{i:\text{n}}}  & \text{if } Z_i = 0 \text{ and } D_i^{\text{obs}} = 0,
	\end{cases}$$

$$w_{i:\text{c}} = \begin{cases} 
	\frac{\widehat{\pi}_{i:\text{c}}}{\widehat{\pi}_{i:\text{c}} + \widehat{\pi}_{i:\text{a}}}  & \text{if } Z_i = 1 \text{ and } D_i^{\text{obs}} = 1 \\
	0 & \text{if } Z_i = 1 \text{ and } D_i^{\text{obs}} = 0 \\
	0 & \text{if } Z_i = 0 \text{ and } D_i^{\text{obs}} = 1 \\
	\frac{\widehat{\pi}_{i:\text{c}}}{\widehat{\pi}_{i:\text{c}} + \widehat{\pi}_{i:\text{n}}}  & \text{if } Z_i = 0 \text{ and } D_i^{\text{obs}} = 0.
	\end{cases}$$

For each of the three strata, calculate the weighted difference in means of the observations.  
The different weights produces the different $ITT$ estimates.

\paragraph{Weak PI and exclusion restriction for Never Takers.}
%
For the treatment side we leverage principal ignorability to estimate $\mu_{\text{c}1}$ and $\mu_{\text{a}1}$. 
We estimate the relevant means for those assigned to treatment via:
\begin{align*}
\widehat{\mu}_{\text{c}1} = \frac{ \sum_{i} \widehat{\phi}_{i} Y_i^\obs  1_{\{Z_i = 1, D_i^\obs = 1\}} }{ \sum_{i}  \widehat{\phi}_{i}1_{\{Z_i = 1, D_i^\obs = 1\}}}; \qquad 
\widehat{\mu}_{\text{a}1} = \frac{ \sum_{i} (1-\widehat{\phi}_{i}) Y_i^\obs 1_{\{Z_i = 1, D_i^\obs = 1\}}}{ \sum_{i}(1 -  \widehat{\phi}_{i})1_{\{Z_i = 1, D_i^\obs = 1\}}},
\end{align*}
with $\widehat{\phi}_{i} = \widehat{\pi}_{i:\text{c}}/(\widehat{\pi}_{i:\text{c}} + \widehat{\pi}_{i:\text{a}})$ being the ratio of principal scores.

For the control side, the exclusion restriction gives
$$\widehat{\mu}_{\text{c}0} = \overline{Y}_{00} \frac{\widehat{\pi}_\text{c} + \widehat{\pi}_{\text{n}} }{\widehat{\pi}_\text{c}} - \overline{Y}_{10} \frac{\widehat{\pi}_{\text{n}}}{\widehat{\pi}_\text{c} }; \qquad \widehat{\mu}_{\text{a}0} = \overline{Y}_{01}.$$

We finally take the differences to obtain the estimates for $\text{ITT}_{\text{c}}$ and $\text{ITT}_{\text{a}}$.

\subsubsection{Using covariates to model outcomes}
The discussion above has focused on the use of covariates for justifying principal ignorability and estimating the principal score. 
In practice, we can also leverage covariates that are predictive of the outcome to sharpen inference for the causal effects themselves.~\citet{Jo:2009hx} propose a straightforward strategy of a weighted regression of $Y^{\obs}$ on $Z$ and $X$ with the relevant principal score weights.~\citet{Ding:2014wc} borrow methods from survey sampling and discuss model-assisted estimation, which reduces to the strategy in~\citet{Jo:2009hx} in certain settings. These approaches are sensible for both one- and two-sided noncompliance, so long as we are only utilizing principal ignorability assumptions. However, neither approach readily extends to the ``mixed'' case of weak PI and the exclusion restriction. We do not explore this question further here, though note that post-stratification is one possible strategy for additional covariate adjustment.

\section{Application to Head Start Impact Study\label{sec:hsis}}
In the Head Start Impact Study, families applied to enroll their eligible children in a given Head Start center. In roughly 350 Head Start centers, the offer of enrollment was randomly assigned among eligible children. 
Of those offered a spot, 18\% of children in our analysis sample were Never Takers who did not actually enroll (i.e., $\widehat{\pi}_n = 0.18$). 
In addition, 13\% of children not offered the opportunity to enroll were Always Takers who nonetheless enrolled in a Head Start center during the study period (i.e., $\widehat{\pi}_a = 0.13$). 
Roughly half of the observed Always Takers enrolled in the center of randomization (i.e., where they were formally denied access to the program for that year) and half enrolled in a different Head Start Center~\citep{puma2010hsis}.
Finally, this leaves $\widehat{\pi}_{\text{c}} = 0.69$ Compliers in the sample.

Since the goal of the study is to estimate the effect of enrolling in Head Start on various outcomes, the standard approach would be to invoke the usual instrumental variables assumptions to estimate $\text{ITT}_{\text{c}}$: monotonicity and the exclusion restrictions for Always Takers and Never Takers~\citep{angrist1996identification}. While monotonicity and the exclusion restriction for Never Takers are both highly plausible in this case, the exclusion restriction for Always Takers is somewhat more controversial. As~\citet{Gibbs:2011uz} argue, if the centers of enrollment for Always Takers systematically differ from their centers of randomization, then the exclusion restriction might not hold for this group~\citep[see also][]{bloom_weiland2014}. Thus, we propose using principal score methods to explore the effect of the exclusion restriction for Always Takers on estimates of $\text{ITT}_{\text{c}}$. Following earlier analyses~\citep{ding2015randomization} and to simplify exposition, we restrict our attention to a complete-case subset of HSIS, with $N_1 = 2,238$ in the treatment group and $N_0 = 1,348$ in the control group. For illustration, our outcome of interest is the Peabody Picture Vocabulary Test (PPVT), a widely used measure of cognitive ability in early childhood. For covariates, we will adopt the rich set of child- and family-level covariates used in the original HSIS analysis of~\citet{puma2010hsis}, including pre-test score, child's age, child's race, mother's education level, and mother's marital status. In total, there are $k = 20$ covariates after re-coding factor variables. Despite these important covariates, principal ignorability assumptions are nonetheless quite heroic in this context.

\begin{figure}[btp]
	\centering
	\includegraphics[scale=0.45]{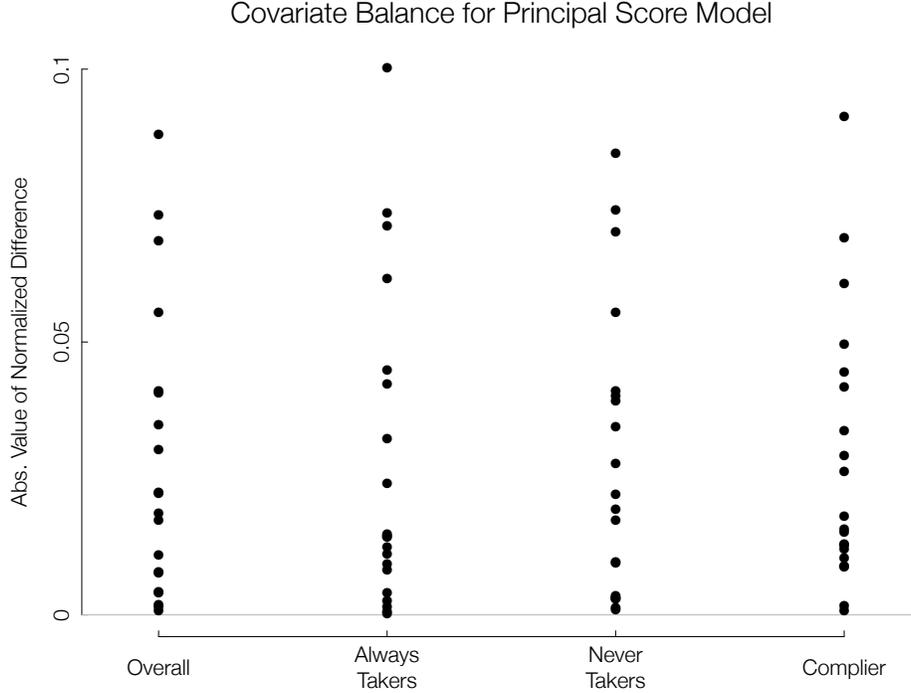}
	\caption{\label{fig:PI_cov_means} Normalized differences by principal stratum for $k = 20$ individual-level covariates, given the principal score model.}
\end{figure}

First, we fit principal score models using the ``marginal method'' in Section~\ref{sec:est_score_two_sided}. That is, we estimate two separate logistic regressions by treatment arm to estimate $\widehat{\pi}_{i:\text{a}}$ and $\widehat{\pi}_{i:\text{n}}$, and then subtract to estimate $\widehat{\pi}_{i:\text{c}} = 1 - \widehat{\pi}_{i:\text{a}} - \widehat{\pi}_{i:\text{n}}$. In this example, we use only main effects for each of $k = 20$ covariates; adding in higher-order interactions gave comparable covariate balance. We then assess covariate balance given the estimated principal score via the \textit{normalized difference} for each child-level covariate within each principal stratum,
$$\widehat{\Delta}_s = \frac{ \overline{X}_{st} - \overline{X}_{sc} }{\sqrt{(s^2_{st} + s^2_{sc})/2}},$$
where the covariate mean and standard deviation, $\overline{X}_{sz}$ and $s^2_{sz}$, are either calculated directly from the observed data or via the weighting method described above~\citep{imbens_rubin2014}. See~\citet{Ding:2014wc} for additional discussion of covariate balance for principal scores. Figure~\ref{fig:PI_cov_means} shows the normalized differences for the Head Start Impact Study given the estimated principal score. All differences are below $0.1$ in absolute value, suggesting that there is good covariate balance given the principal score. 
We also estimated the principal score via the ``joint method,'' (not shown) which restricts $\widehat{\pi}_{i:\text{c}} \in [0,1]$; this yielded nearly identical results.

\begin{figure}[btp]
	\centering
	\includegraphics[width=\textwidth]{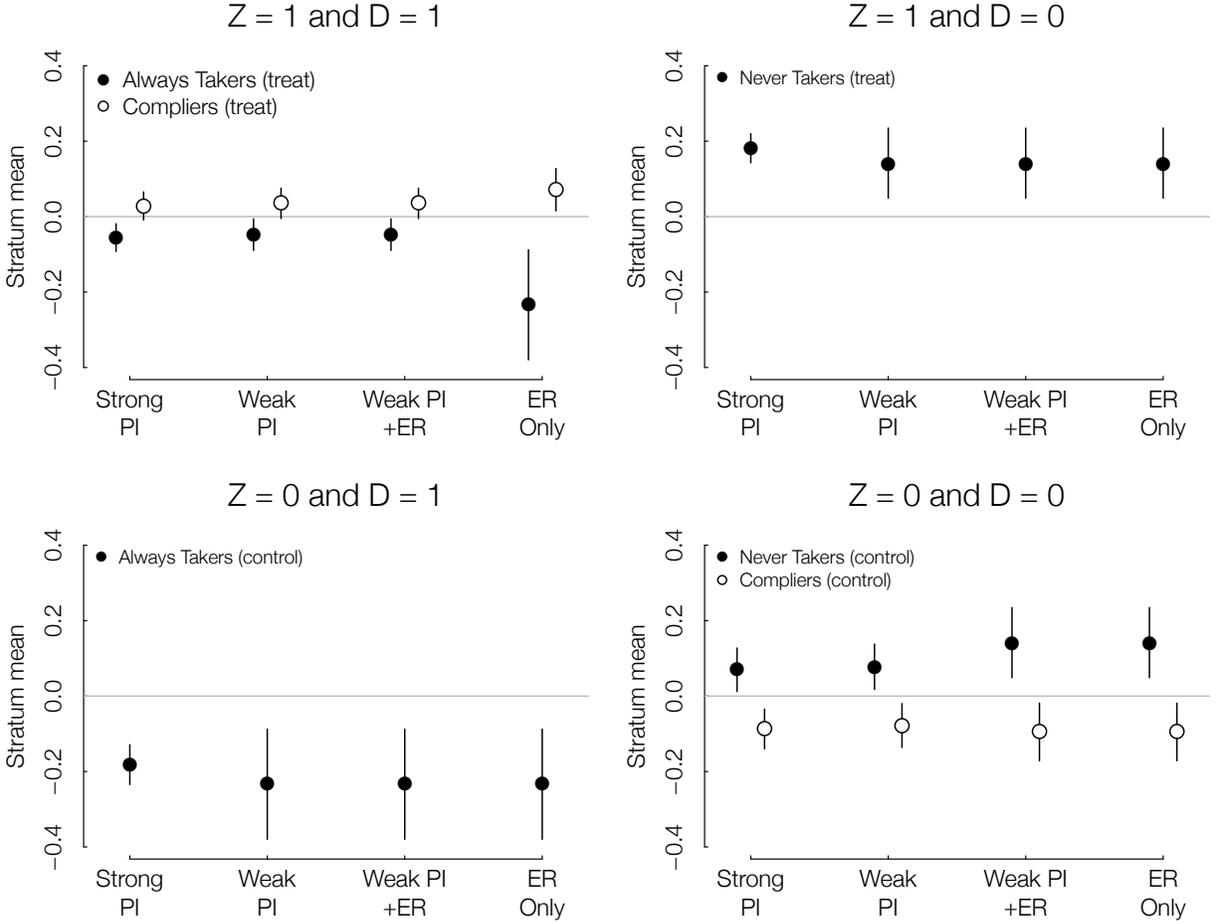}
	\caption{\label{fig:PI_means} Means by principal stratum under different assumptions.}
\end{figure}

\begin{figure}[btp]
	\centering
	\includegraphics[width=\textwidth]{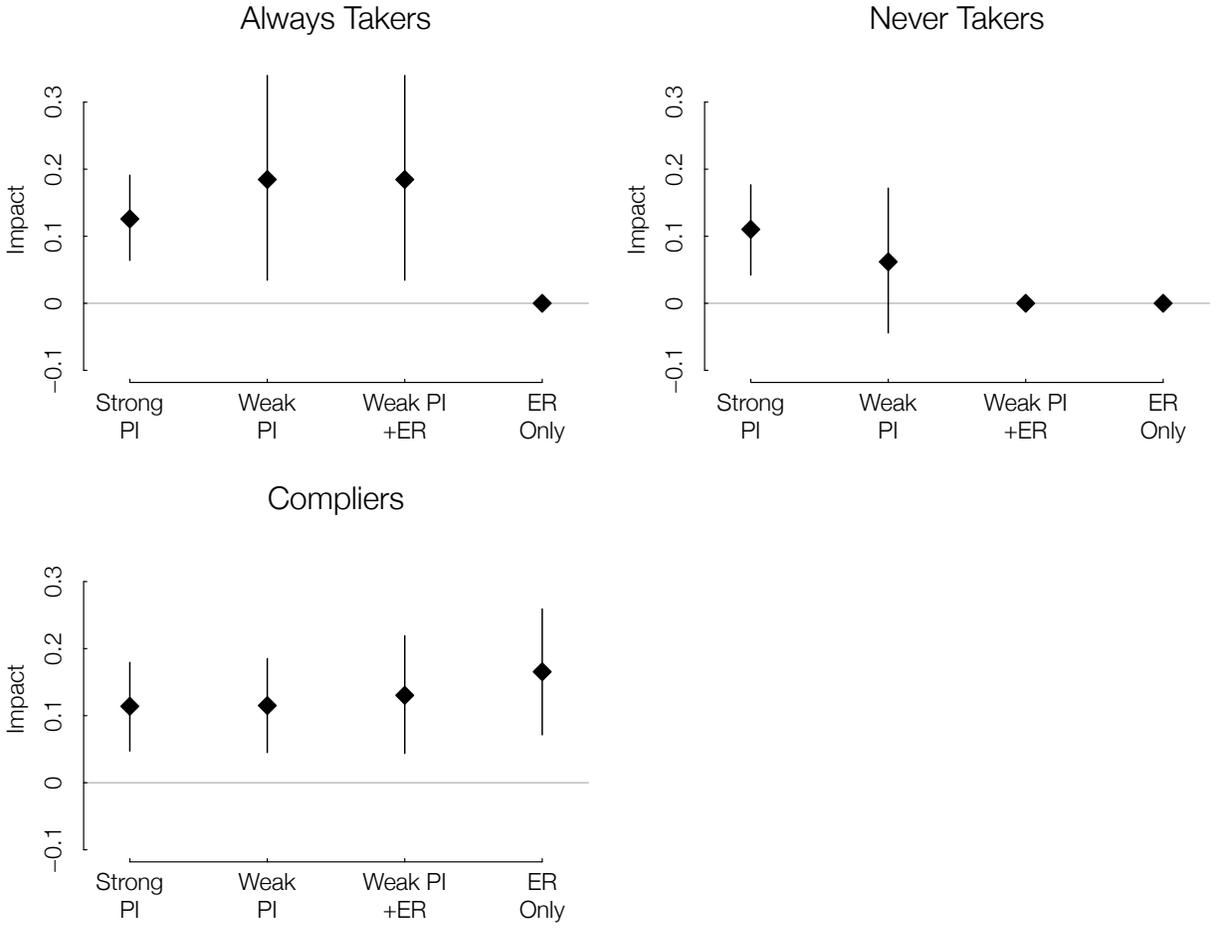}
	\caption{\label{fig:PI_impacts} Impacts by principal stratum under different assumptions.}
\end{figure}

We then estimated principal causal effects under our different assumptions to see how our estimates changed. 
Figures~\ref{fig:PI_means} and~\ref{fig:PI_impacts} show the estimated principal stratum means and impacts, respectively, given (1) Strong Principal Ignorability; (2) Weak Principal Ignorability; (3) Weak Principal Ignorability plus the exclusion restriction for the Never Takers; (4) Exclusion restrictions for both Always Takers and Never Takers.

We compare invoking the exclusion restriction for Never Takers vs an ignorability assumption. As shown in the bottom-right panel of Figure~\ref{fig:PI_means}, the estimate for $\mu_{\text{n}0}$ changes very little with and without the exclusion restriction. Figure~\ref{fig:PI_impacts} shows the same change in terms of impacts, which emphasizes that the estimate for $\text{ITT}_{\text{n}}$ under Weak PI is not meaningfully different from zero (i.e., the exclusion restriction). Thus, estimates assuming weak PI for Compliers and Never Takers do not yield any evidence against the exclusion restriction for Never Takers.

By contrast, consider the exclusion restriction for Always Takers. 
As shown in the top-left panel of Figure~\ref{fig:PI_means}, the estimate for $\mu_{\text{a}1}$ under weak PI is quite different from under an exclusion restriction. 
Figure~\ref{fig:PI_impacts} displays the same change in terms of impacts. 
While estimates for $\text{ITT}_{\text{a}}$ are highly uncertain, they are nonetheless consistently positive and away from zero. 
This result suggests that, based on observable characteristics alone, we should be wary of the exclusion restriction for Always Takers in HSIS.
In the end, however, the estimates for $\text{ITT}_{\text{c}}$ across these different assumptions all are quite similar, as shown in Figure~\ref{fig:PI_impacts}. 


We also considere the testable implications of Strong PI in this example. The top-right and bottom-left panels of Figure~\ref{fig:PI_means} show the estimates for $\mu_{\text{a}0}$ and $\mu_{\text{n}1}$ respectively, the two principal stratum means that we can directly observe in this example. Since estimates are largely unchanged under Strong PI and Weak PI, we do not find evidence against Strong PI in this case. 
Since Weak PI is the strictly weaker assumption, we would therefore prefer that in practice.

\section{Discussion\label{sec:discussion}}
While principal score methods are gaining popularity in the social sciences especially, these methods remain poorly understood. In this paper, we reviewed the literature on principal score methods, highlighted connections between different approaches and applications, and suggested some modest extensions to more complex examples. Finally, we applied this approach to an example from the Head Start Impact Study, finding mixed results. Overall, we argue that researchers should carefully examine the relevant assumptions when using principal score methods, as they can be quite strong in practice.

We briefly discuss several directions for future work. First, while we only discussed simple estimators in the main text, principal score models can be quite rich.~\citet{feller_dissertation}, for example, estimates the principal score for the Head Start Impact Study using a Bayesian hierarchical model that accounts for the multilevel structure in the experiment. While this necessarily requires additional assumptions for valid inference, this is a promising approach to leverage complex models for otherwise simple analyses.

Another critical direction for future work is sensitivity analysis.
\citet{Ding:2014wc} take an important step in this direction, proposing formal sensitivity analyses analogous to approaches for observational studies first introduced by~\citet{rosenbaum1983assessing}. This is especially important because, as discussed above, Principal Ignorability assumptions are quite strong. 
One potentially fruitful approach---essentially a quick-and-dirty sensitivity analysis---is to compare principal score estimates and their corresponding nonparametric bounds. 
The bounds give a range of plausible parameter values, and the principal score estimate gives a ``reasonable guess'' within this interval as to where the truth might be within those bounds. Furthermore, a principal score estimate outside these bounds---though unlikely to occur in practice---would be strong evidence against principal ignorability.
Given available tools, this is an attractive approach.
 
Finally, principal scores are useful objects for describing trends in data even in the absence of Principal Ignorability assumptions, just as the propensity score can be useful in settings other than observational studies. 
In particular, they can be used to describe trends in how individuals respond to the offer of treatment, which is often of substantive interest in its own right.
We are also currently exploring how, even without the ignorability assumptions, principal scores can be used to tighten nonparametric bounds~\citep[see also][]{Long:2013hs}. We anticipate that there will be many other uses.

\clearpage
\bibliographystyle{chicago}
\bibliography{references}

\clearpage
\appendix
\section{Appendix\label{sec:pscore_proofs}}

This proof are nearly identical to the analogous proofs for the propensity score in~\citet{imbens_rubin2014}. Following that example, we first show that the principal score is indeed a balancing score. For convenience, let $C_i$ be an indicator for whether student $i$ is a Complier. For this we need
$$C_i \ind \mathbf{X}_i \mid \pi_i$$
to hold, or equivalently:
$$\mathbb{P}\{C_i = 1 \mid \mathbf{X}_i, \pi_i\} = \mathbb{P}\{C_i = 1 \mid \pi_i\}.$$
We will show that both sides of the equation equal $\pi_i$. For the left hand side, $\mathbb{P}\{C_i = 1 \mid \mathbf{X}_i, \pi_i\} = \mathbb{P}\{C_i = 1 \mid \mathbf{X}_i\} = \pi_i$. For the right hand side:
$$\mathbb{P}\{C_i = 1 \mid \pi_i\} = \mathbb{E}\{C_i \mid \pi_i\} = \mathbb{E}\{\mathbb{E}\{C_i \mid \mathbf{X}_i, \pi_i\} \mid \pi_i \} = \mathbb{E}\{\pi_i \mid \pi_i \} = \pi_i.$$
Therefore the principal score is a balancing score.

Second, we show that if Strong Principal Ignorability holds given $\mathbf{X}_i$, Strong Principal Ignorability also holds given $\pi_i$. We show this for $Y_i(0)$, with an identical argument for $Y_i(1)$. Therefore, we need to show that:
$$Y_i(0) \ind C_i \mid \pi_i$$
holds, or equivalently:
$$\mathbb{P}\{C_i = 1 \mid Y_i(0), \pi_i\} = \mathbb{P}\{C_i = 1 \mid \pi_i\}.$$

To show this:
\begin{align*}
\mathbb{P}\{C_i = 1 \mid Y_i(0), \pi_i\} &= \mathbb{E}\{C_i \mid Y_i(0), \pi_i\} \\
&= \mathbb{E}\left\{\mathbb{E}\{C_i \mid Y_i(0), \mathbf{X}_i, \pi_i\} \mid Y_i(0), \pi_i \right\} \\
&= \mathbb{E}\left\{\mathbb{E}\{C_i \mid \pi_i\} \mid Y_i(0), \pi_i \right\} \\
&= \mathbb{E}\{C_i \mid \pi_i\} = \mathbb{P}\{C_i = 1 \mid \pi_i\}\\
\end{align*}
where we use Principal Ignorability and the fact that the principal score is a balancing score to go from the second to third lines. Therefore, Strong Principal Ignorability also holds given $\pi_i$.

\end{document}